# Language-native reasoning over lightly structured knowledge for materials synthesis planning

*Yuze Liu,[1, 3†] Yihe Zhang,[2†] Zhaoyuan Zhang,[1,3] Xiangsheng Zeng,[2] Leping Yu,[2] Liu Yang,[4] Lejia Wang,[2, 5*] Xi Yu[1*]*

1. *State Key Laboratory of Advanced Materials for Intelligent Sensing, Key Laboratory of Organic Integrated Circuit, Ministry of Education & Tianjin Key Laboratory of Molecular Optoelectronic Sciences, Department of Chemistry, School of Science, Tianjin University, Tianjin 300072, China*
2. *Boronmatrix Advanced Materials Techology Co., Ltd.*, *No 3938 Wenchuan Road, Baoshan District, Shanghai City, 200949, China*
3. *Language Intelligence Technology Co., Ltd., Haihe Education Park, Jinnan District, Tianjin City, 300350, China*
4. *College of Intelligence and Computing, Tianjin University, Tianjin 300072, China*
5. *School of Materials and Chemical Engineering, Ningbo University of Technology, Ningbo, 315211, China*

** Authors to whom correspondence should be addressed: xi.yu@tju.edu.cn*, or *wanglejia@nbut.edu.cn*

**Abstract:** Materials synthesis procedures are predominantly documented as narrative text in papers, protocols, and laboratory records, placing them beyond the reach of conventional data-driven optimization frameworks. This language-native character poses a particular challenge for complex, multistage processes such as the preparation of boron nitride nanosheets (BNNS), where outcomes depend on path-dependent choices in exfoliation, functionalization, and functionalization. Here, we recast synthesis planning of the materials as a text reasoning problem enabled by a lightly structured knowledge substrate that preserves the procedural logic and causal contexts while exposing computable elements for retrieval. Built on this representation, our framework combines semantic matching, lexical search, and parameter-aware filtering to support retrieval-augmented generation with more accurate and better-grounded synthesis guidance. We further introduce experience-augmented reasoning, in which iteratively refined text guides distilled from multi-source narratives support hypothesis generation, failure diagnosis, and protocol revision. We validated the framework in the targeted exfoliation of BNNS, a synthesis problem governed by multivariate constraints and limited transferability of literature protocols across laboratory settings. By integrating dispersed literature evidence with experimentally observed failure modes, the system converged within only three iterative rounds on a high-performing protocol that yielded high-quality ultrathin nanosheets meeting the target specifications, substantially shortening what is often a prolonged cycle of expert-led trial-and-error. By enabling language-native reasoning over procedural knowledge, this framework moves AI beyond literature assistance toward active synthesis planning, adaptation and acceleration in complex materials workflows.

## Introduction

Artificial intelligence is reshaping materials research,[1-4] yet its role in synthesis planning remains far less developed than in property prediction and materials screening. [5-8] In material synthesis and processing, critical knowledge is often "language-native," residing in papers, protocols and laboratory records as procedural narrative rather than tabulated variables. [9,10] Experimental success depends not only on the values of individual parameters, but also on the order of operations, conditional dependencies, trade-offs and failure contingencies encoded in text. As a result, many experimentally important decisions remain difficult to formalize within machine-learning frameworks that rely primarily on structured data representations.[11]

The preparation of boron nitride nanosheet (BNNS) serves as a compelling example of this challenge,[12,13] where its preparation involves a tightly coupled, multistage workflow spanning pretreatment, exfoliation and post-processing and the upstream choices and downstream processing matches. Experimentalists must navigate a path-dependent decision space[14], that are documented almost exclusively in textual literature and lab notes. BNNS preparation, therefore, constitutes a multifactorial decision problem that requires interpreting procedural logic rather than merely optimizing numerical variables, calling for a system capable of reasoning over textual knowledge in a form relevant to experimental decision-making.[15]

Large Language Models (LLMs) offer a promising route for working with such language-native evidence,[16-19] because of their inherent capacity to process procedural text. However, current implementations in materials science generally treat LLMs as extraction tools or interfaces to rigid database, [20-25] leaving their role in synthesis planning still limited. Information extraction pipelines,[26-30] for instance, effectively organize entities into knowledge graphs[31-37] but often flatten procedural nuances — stripping away the dynamic conditionals, feedback loops, and causal links that are essential for decision-making. Similarly, although emerging autonomous agents[38,39] have shown promise in protocol execution[40-45] and parameter tuning[46], they remain less effective in situations that require weighing qualitative trade-offs across multistage workflow. In practice, synthesis planning is a dependency-rich decision problem that requires reasoning across contingencies rather than simply retrieving facts.[47] Advanced LLMs [48] are therefore promising not only as tools for information access, but also as systems that can operate over conditional logic, path branching and procedural dependencies embedded in textual narratives. When coupled to a representation that preserves such context, they can support deliberation over multivariate constraints in flexible, path-dependent materials processing workflows.[16]

In this study, we introduce a Language-Native Materials Processing (LNMP) framework that treats natural-language experimental evidence as a primary data modality for synthesis design. As illustrated in **Fig. 1**, we construct a heterogeneous database constructed from more than 3,500 BNNS-related publications using modular LLM prompts (see the Preparation module example in **Fig. 3a**). Full-article content—including preparation protocols, characterization results, mechanistic inferences, theoretical modeling, and tabular data—is converted into lightly structured records. These records comprise provenance-linked narrative modules together with typed fields typed fields for computable elements such as conditions, materials and outcomes. In this way, the light structuring provides a stable organizational backbone while preserving the contextual richness of the original narratives under higher-level headings such as preparation, characterization, mechanisms, modeling, and retaining verifiable, evidence-linked snippets at the subheading level. Unlike rigid schemas that discard context, this design organizes experimental knowledge into semantic modules for targeted reasoning while retaining computable entities and relations for precise filtering. In this way, the framework preserves the procedural logic of synthesis narratives while retaining computable elements for precise filtering, thereby supporting hybrid retrieval across semantic, lexical and parametric dimensions.

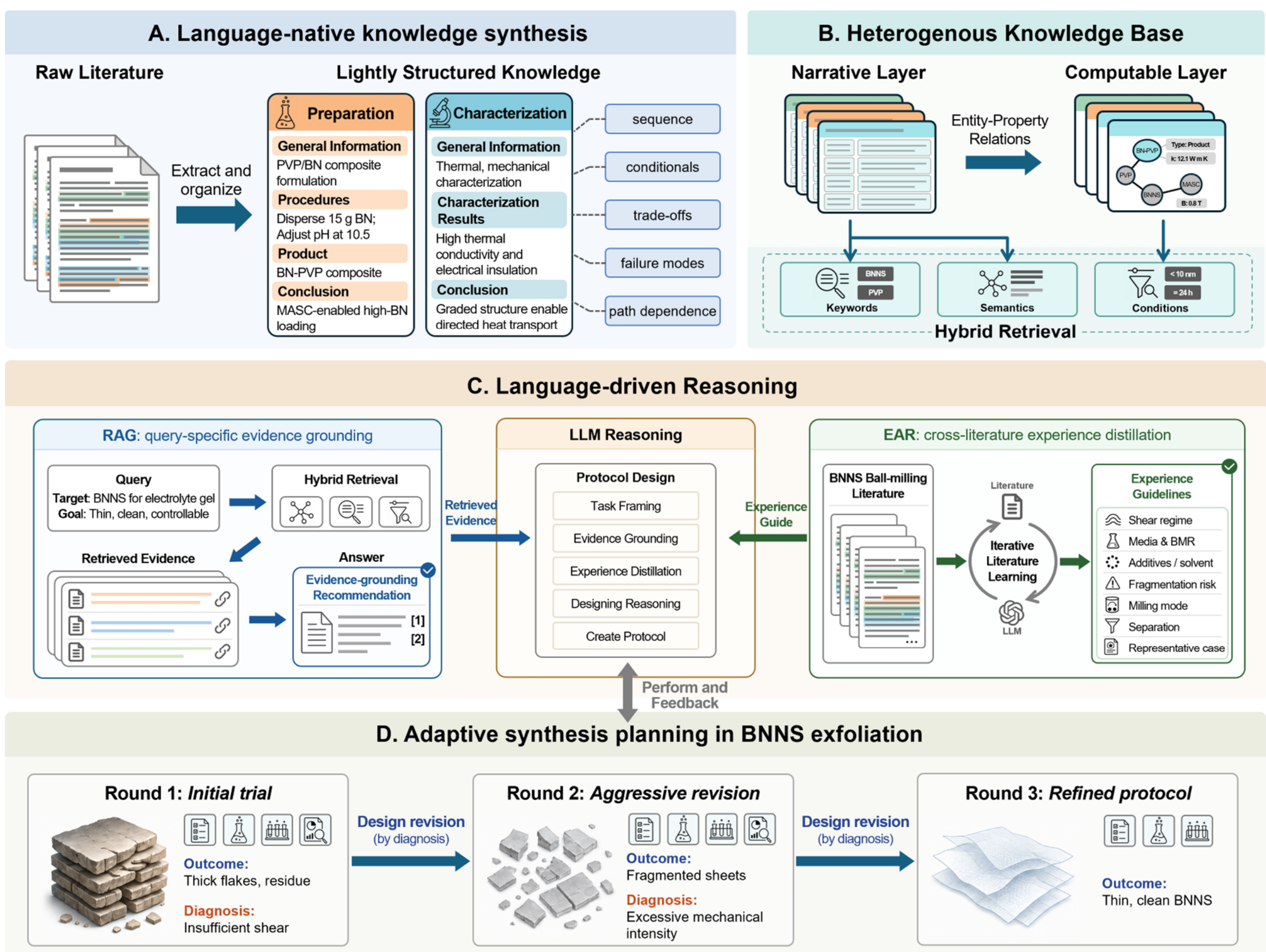

**Figure 1:** Language-native AI framework for BNNS materials processing design. (a) Scientific literature is transformed into lightly structured, provenance-linked text records that retain experimental context. (b) A narrative layer and a computable layer jointly organize the lightly structured records, enabling hybrid retrieval from the database. (c) Retrieval-augmented generation (RAG) and experience-augmented reasoning (EAR) convert retrieved evidence and distilled experimental experience into iterative synthesis guidance. (d) The framework is applied to BNNS exfoliation to guide experimental design and protocol refinement.

Built on this lightly structured database, literature-derived experimental knowledge can first be organized into traceable evidence for retrieval-grounded recommendation. In this mode, retrieval-augmented generation (RAG) grounds recommendations in specific literature modules and enables more accurate, complete and actionable responses to complex materials queries, particularly when useful evidence is dispersed across multiple studies. Beyond this evidence-grounded use, the same database also supports Experience-Augmented Reasoning (EAR), in which multi-source narratives are distilled into reusable "experience units" that capture heuristics, failure modes and laboratory priors for iterative planning, failure diagnosis and protocol revision. We applied this capability to the targeted exfoliation of BNNS, a process governed by strong path dependence and multivariate constraints. By reasoning over the hybrid database and adapting to experimentally observed outcomes, the system identified effective combinations of grinding additives, milling configurations and separation strategies within three iterative rounds. These AI-guided protocols were experimentally verified to yield high-quality BNNS, demonstrating that integrating lightly structured knowledge with LLM-driven reasoning can support adaptive, mechanism-guided process optimization beyond literature recommendation alone.

## Encoding Materials Knowledge via Lightly Structured Databases

### Narrative and Lightly Structured Dataset

The dataset for BNNS was derived from over 3,500 literature articles identified through a systematic search and screening process (**Table S1**; **Supplementary Note 1**). To transform this unstructured corpus, we applied a modular LLM prompting strategy that generates "lightly structured" narrative data (see Preparation module example in **Fig. 2a** and full prompt templates used for data extraction are provided in **Supplementary Data 2**). The prompt format imposes a hierarchical structure where high-level semantic categories (e.g., preparation, characterization, and mechanism) form a stable backbone, while keep the lower-level subfields flexible to preserve evidence-linked narrative snippets (Fig. 2a). This representation effectively balances machine parsability with semantic richness by avoiding the over-compression of standard extraction pipelines[49-51], while maintains contextual fidelity by allowing flexible narrative entry under structured headers, ensuring that the full provenance of the synthesis process—including path-dependent logic—is preserved for downstream reasoning.

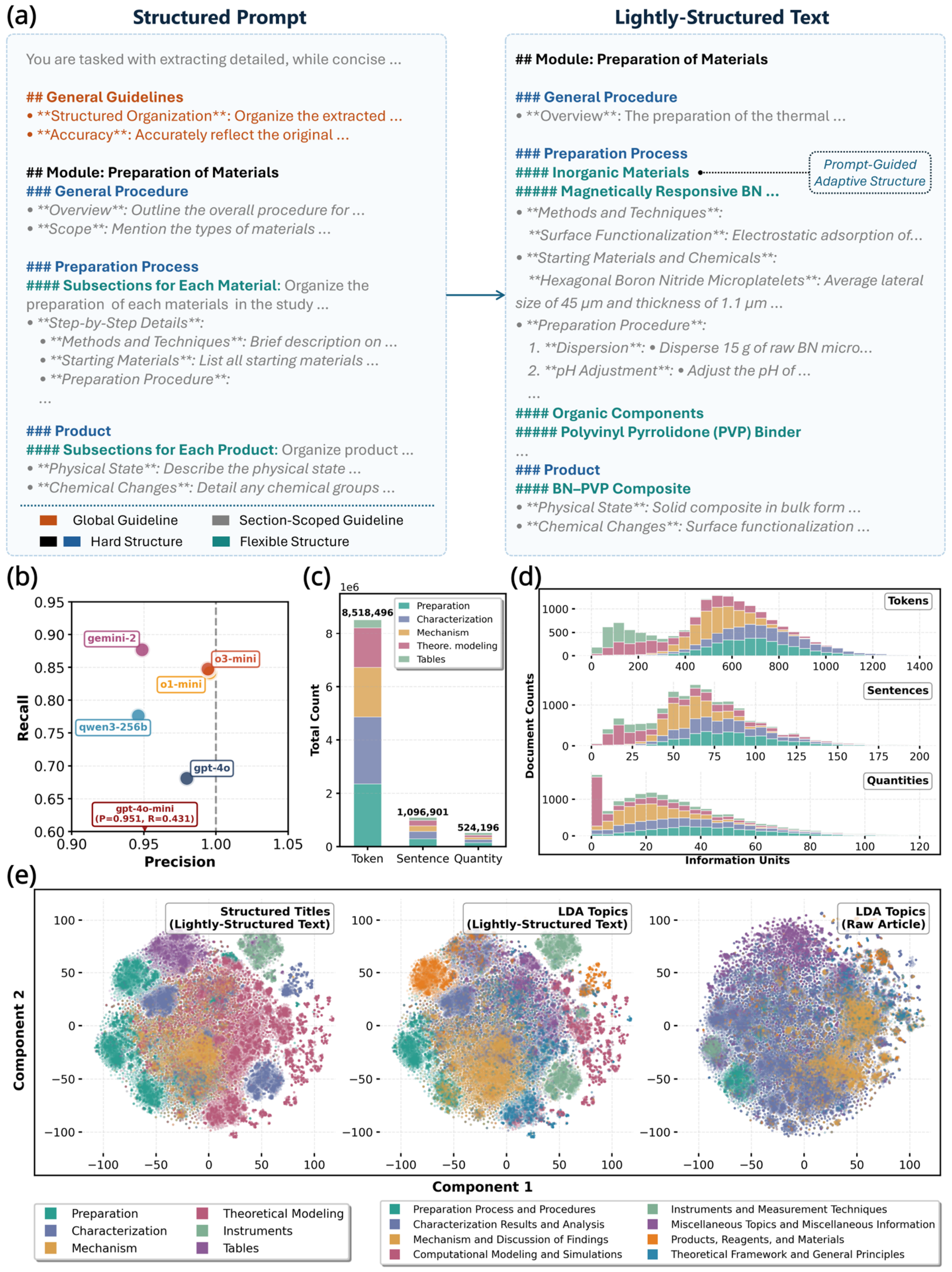

**Figure 2:** Extraction and analysis of lightly structured texts. (a) Illustrates the construction of structured prompts by integrating hard and flexible structural headings, guided by domain-specific knowledge, to enable LLMs to generate lightly structured text. (b) Shows the performance of different large language models in precision and recall when mapping numerical information from 80 sample articles (c-d) Distribution of information units in the extracted lightly structured texts, with (d) illustrating the variation in token, sentence and quantity counts across documents. (e) t-SNE visualization of the lightly structured text, comparing customized title (left 1) with LDA-derived topic (left 2); the strong alignment between title boundaries and topic distributions indicates that the lightly structured title capture coherent semantic module, whereas LDA of original texts (right 1) reveals more topic mixing, highlighting the enhanced focus achieved through reconstruction.

We evaluated the fidelity and structure of this dataset through quantitative benchmarks and topological analysis (**Fig. 2b-e**; detailed extraction metrics and compression ratios are shown in **Supplementary Figs. S2-5**). To assess extraction reliability, we mapped numerical information from 80 sampled articles into light-structured formats using various LLMs. (**Fig. 2b;** comparative extraction outputs for all tested models are listed in **Supplementary Data 3**). Reasoning-capable models such as o1-mini and o3-mini significantly outperformed standard models, with o3-mini achieving the highest recall (~0.85). Notably, while models like gemini-2.5 achieved high recall, they exhibited lower precision due to hallucinations, effectively prioritizing structural coherence over factual fidelity by generating plausible but unsupported data to fill schema gaps. These results indicate that reconstructing scientific text into lightly structured representations requires not only pattern recognition, but also reasoning over the contextual dependencies embedded in procedural narratives.

The resulting corpus spans ~8.5 million tokens with a content distribution where "Preparation" and "Characterization" modules dominate the procedural knowledge base (**Fig. 2c-d**). The structural advantage of this organization is further visualized via t-SNE[52] and LDA analysis[53] (**Fig. 2e**). Compared to raw text, the lightly structured topics form significantly tighter clusters with reduced topic mixing. This improvement is particularly important since in raw text, preparation procedures, characterization data, mechanistic interpretation and theoretical analysis are often interwoven within the same raw text, increasing the difficulty of both effective retrieval and downstream LLM reasoning. By disentangling these heterogeneous information types into more coherent, task-relevant modules, the lightly structured organization provides a more suitable basis for evidence retrieval and reasoning.

### Dual-Layer Organization of Narrative and Computable Data

Building upon the extracted narratives, we constructed a heterogeneous database schema designed to bridge the gap between semantic context and precise data querying **(Fig. 3a)**. This architecture transforms raw articles into a multi-layered system underpinned by robust external indexing. At one level, the Narrative Layer preserves article metadata (e.g., IDs, titles, authors) together with lightly structured modules that organize content into semantic categories such as Preparation and Characterization. These modules further incorporate subfields such as Process, Product, Instrument and Results, reflecting the input–output logic commonly used in materials research.

Complementing the narrative backbone is the Computable Structure Layer, which translates prose into machine-actionable filters precise by computable entities via Named Entity Recognition (NER)[26] and relational dependencies through Knowledge Graphs (KG).[37] It enables the system to traverse relationships and enforce rigorous constraints—such as specific parameter ranges (e.g., thickness < 100 nm) or processing times—that are impossible to isolate through semantic search alone. To operationalize this dual-layer data, we implemented a Hybrid Indexing Mechanism that fuses semantic breadth with lexical precision. Dense vector indices (utilizing BGE-M3 models[54] with FAISS L2 indexing[55]) capture conceptual similarities, while inverted indexes utilizing BM25 ensure exact keyword matching. This combination empowers the engine to balance the "fuzzy" conceptual understanding of vector retrieval with the "hard" constraints of KG parameters.[56-58] By integrating structural relational constraints with semantic narrative retrieval, this architecture significantly reduces information redundancy while ensuring that retrieved evidence is both textually relevant and parametrically valid.

### Context-Aware Retrieval for Generative Reasoning

**Fig. 3b** illustrates the end-to-end hybrid retrieval pipeline. To avoid the context fragmentation typical of standard chunk-based indexing,[50,51,59-61] our system enforces a strategy of module-level retrieval. By leveraging the lightly structured narrative layer, the system retrieves complete semantic units (e.g., *Preparation*, *Mechanism*) rather than isolated text segments, ensuring that the retrieved evidence retains its internal logic and narrative completeness. The process begins with query structuring, where user inputs are decomposed into semantic vectors, lexical keywords, and strict parametric constraints (e.g., lateral size < 500 nm). This drives a multi-axis retrieval engine that filters literature through three simultaneous lenses: BM25 for keyword relevance, embeddings for conceptual matching, and NER for numerical validity. Only modules that satisfy this tripartite filtering are advanced to the ranking stage. Finally, the curated evidence is assembled into a "reasoning-aware" prompt (**Supplementary Data 4**). This structure provides the LLM with a coherent, evidence-backed narrative foundation, enabling it to generate verifiable SOPs and address complex mechanistic challenges—such as interfacial thermal resistance—without the noise typical of unstructured search.

## Benchmarking Reasoning Capabilities

### Surpassing Conventional Retrieval in Scientific Contexts

The efficacy of our system arises from the integration of a heterogeneous database with a hybrid retrieval mechanism. As shown in **Fig. 4a**, we evaluated performance on 39 domain-specific queries (e.g., targeting specific BNNS dimensions or thermal enhancement mechanisms). Our framework achieved a first-hit success rate of 56.4%, outperforming the baseline RAG (33.3%), while failed hits decreased from 38.5% to 5.1%, suggesting substantially improved coverage. **Fig. 4b** further demonstrates that a weighted hybrid strategy (0.6 semantic + 0.3 BM25 + 0.1 KG weights) maximizes relevant matches; detailed ablation results are provided in **Supplementary Figs. S6** and **S7**.

This improvement is primarily attributed to light structural alignment of the data. Unlike conventional RAG that fragment text into arbitrary chunks, our approach organizes lightly structured modules (e.g., Preparation, Characterization) as coherent retrieval units. This design aligns queries with narrative modules, enabling relevance scores to be aggregated at the module level rather than diluted across fragmented segments.

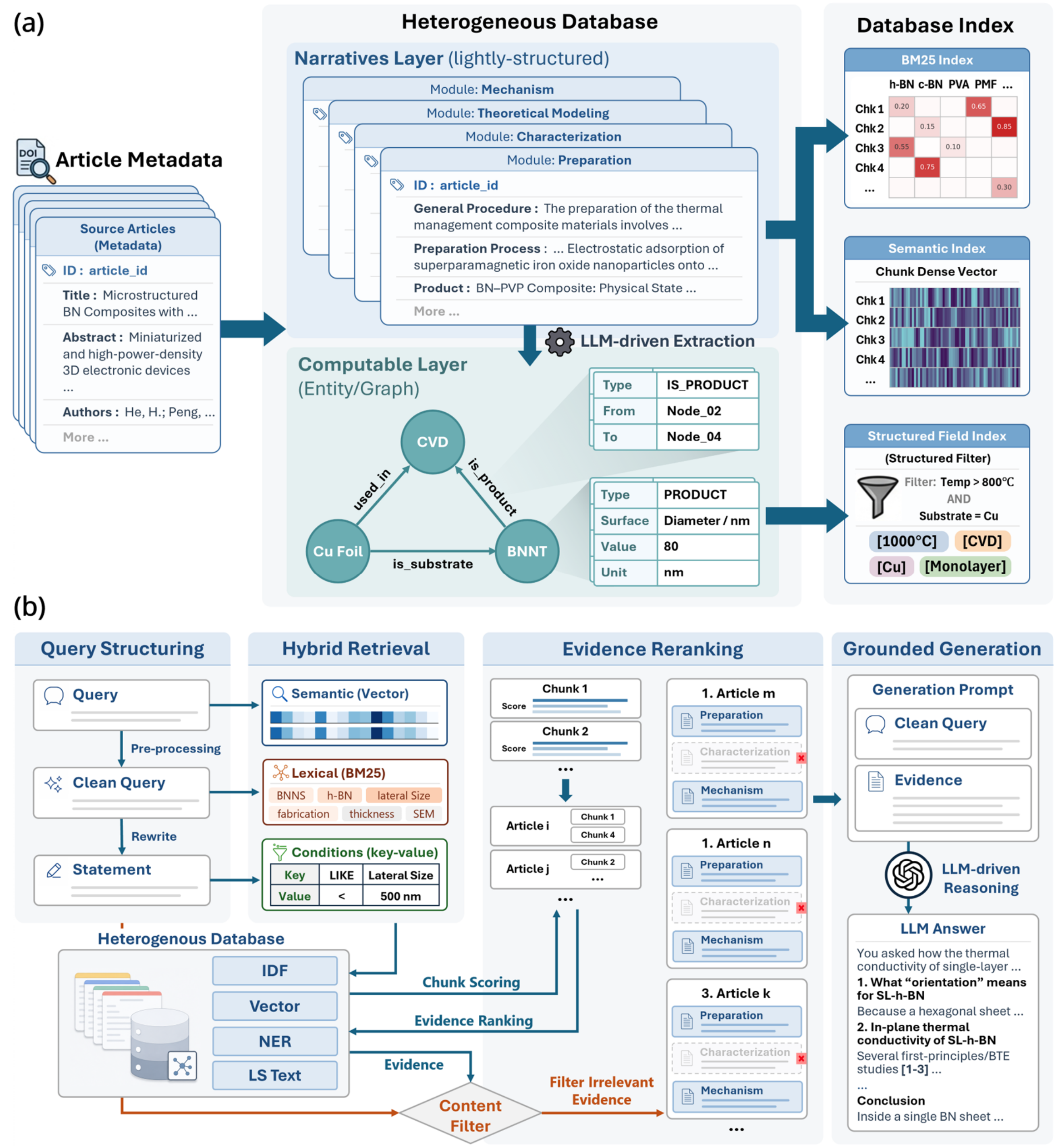

**Figure 3**: Schematic illustration of the heterogeneous database and the RAG architecture. (a) Experimental knowledge is organized into a narrative layer of lightly structured text and a computable layer of structured data. Retrieval is supported by three indices: a lexical index, a BM25-based index, and a structured-field index for constraint-based filtering. (b) The RAG system built on this database operates in four stages: query structuring, hybrid retrieval, evidence re-ranking, and grounded generation, which together transform user queries into factually consistent, evidence-backed outputs.

By providing a high-fidelity evidence substrate, the structured retrieval further enables the LLM

to generate accurate, quantifiable responses that with explicit citations and multi-faceted reasoning, approximating expert-level analysis.

To validate the framework's versatility, we enlisted human experts in BNNS materials studies to curate 60 representative queries spanning five categories: data-retrieval (e.g., extracting specific measurements), recommendation (e.g., suggesting preparation protocols), informational (e.g., explaining mechanisms), integrative summary (e.g., synthesizing trends across studies), and open-ended (e.g., hypothesizing novel composites) tasks. Expert evaluators scored responses on a 1-5 scale for accuracy, completeness, and relevance, comparing our RAG system against: (1) Perplexity (o3 model) with web search capabilities; (2) a standard online model (o3, with external tool-access via OpenAI responses API); and (3) a standard RAG model[60,62] (Chunk Retrieval with o3 model). The complete benchmarking dataset, including the specific queries, model responses, and expert scores rubrics, is provided in **Supplementary Data 5**.

**Fig. 4c** demonstrates the consistent superiority (scores > 4.0) of our framework, particularly in tasks of integrative summaries and recommendations, where aggregating dispersed evidence is essential for multi-hop reasoning. This holistic competence is further captured in **Fig. 4d**, where our system maintains proficiency scores (between 4 and 5) across all axes, clearly outperforming the baselines. While the baseline models (e.g. Perplexity) operated on broader, uncurated search spaces, the superior performance of our framework suggests that domain-specific data curation combined with light-structuring is critical. These results validate the framework's efficacy in reconciling language-native knowledge with computable elements, specifically for material research scenarios dominated by unstructured linguistic data.

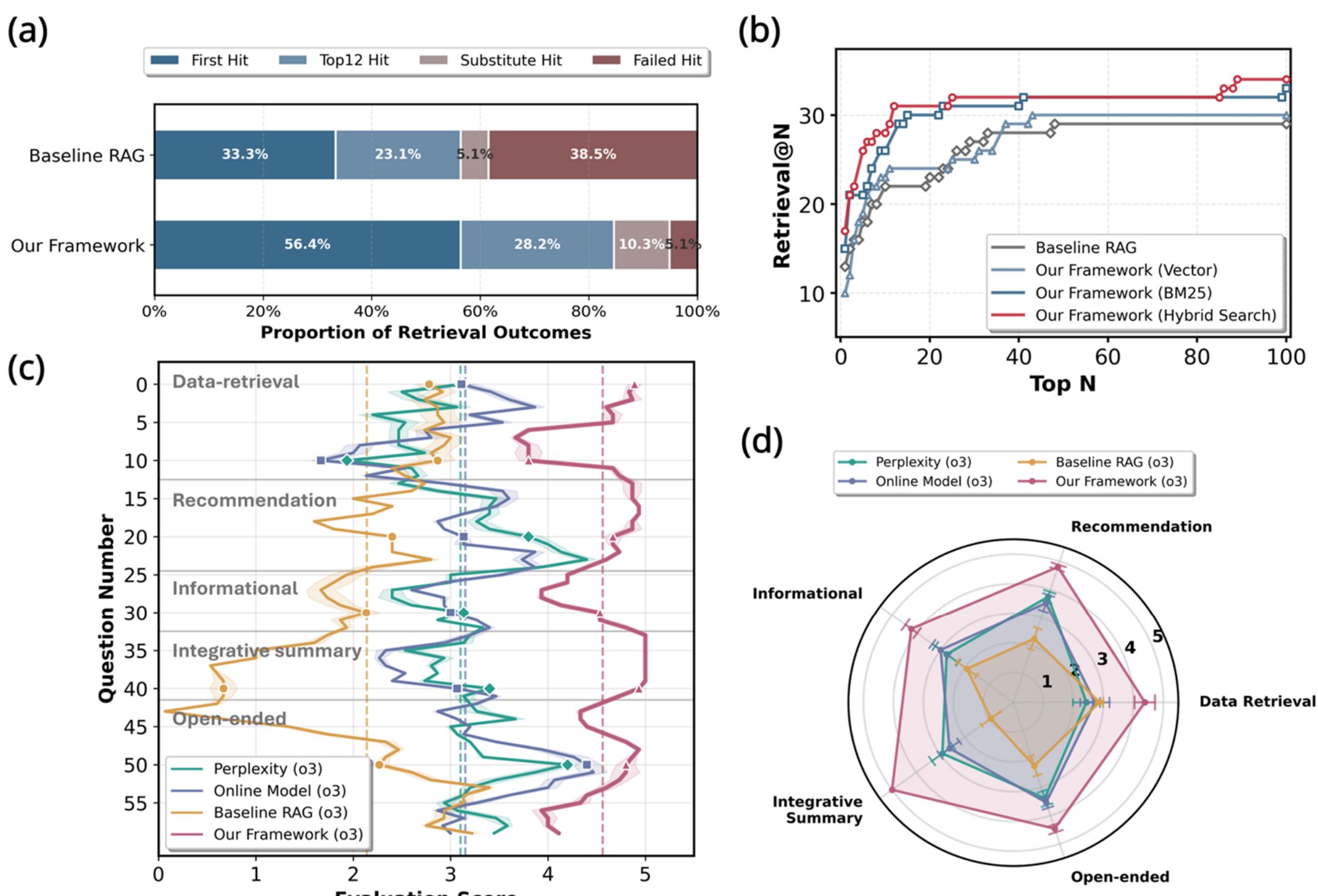

**Figure 4:** Evaluation of retrieval-augmented generation (RAG) performance. (a) Retrieval effectiveness comparison between the baseline RAG system and our proposed framework. (b) Comparison of Retrieval@N across four retrieval strategies: our hybrid framework, keyword-only retrieval, vector-only retrieval, and the baseline RAG system. (c-d) Expert-based manual assessment of 60 domain-specific questions in the BNNS-polymer composites domain, evaluating the factual accuracy and practical utility of generated responses. Methods compared include Perplexity (green), ChatGPT Online (blue), baseline RAG (yellow), and our framework (red), all using o3 as the agentic LLM.

**Expert-Level Resolution of Complex Material Queries**

To provide a granular understanding beyond the aggregate scores illustrated in **Fig. 4c-d**, we analyze three representative cases that span the spectrum of query complexity—from factual retrieval to expert-level reasoning—as detailed in **Fig. 5**. These cases illustrate how the organization of language-native knowledge critically shapes response quality. For brevity, we compare our framework against Perplexity (o3, web-enabled) and a baseline RAG system that employs conventional chunk-based document segmentation; all other components (retriever, generator, prompt template) are held constant. Full query-response transcripts and additional analyses for these and other categories are provided in **Supplementary Note 2** and **Supplementary Data 5**.

**Case 1 (Data Retrieval):** The query requests precise estimation of the thermal conductivity enhancement of a 9:1 MWCNT/BNNS composite relative to a BNNS/PVA baseline. Our system applies structured filtering to match material ratios and experimental setting, retrieves the exact study, extracts the relevant measurements, and computes the differential accurately. In comparison, Perplexity retrieves an unrelated CNF/BNNS study and combines values across independent sources, leading to a source-conflation error. Although the baseline RAG identifies the correct document, it misaligns the condition with values from a different sample group, resulting in numerical error.

**Case 2 (Recommendation):** The query seeks an appropriate mesh cutoff for interfacial thermal modeling, which depends on the trade-off between computational cost and accuracy. Our system synthesizes evidence across multiple studies and derives a transferable recommendation. Begin with a moderate baseline (150 Ry) and increase to 300-800 Ry for defect-sensitive calculations while maintaining grid density to avoid aliasing artifacts. In contrast, both Perplexity and the baseline RAG rely on single-document evidence and simply reproduce the parameter reported in the retrieved paper. This lack of cross-study synthesis yields recommendations that are not generalizable to varying simulation constraints.

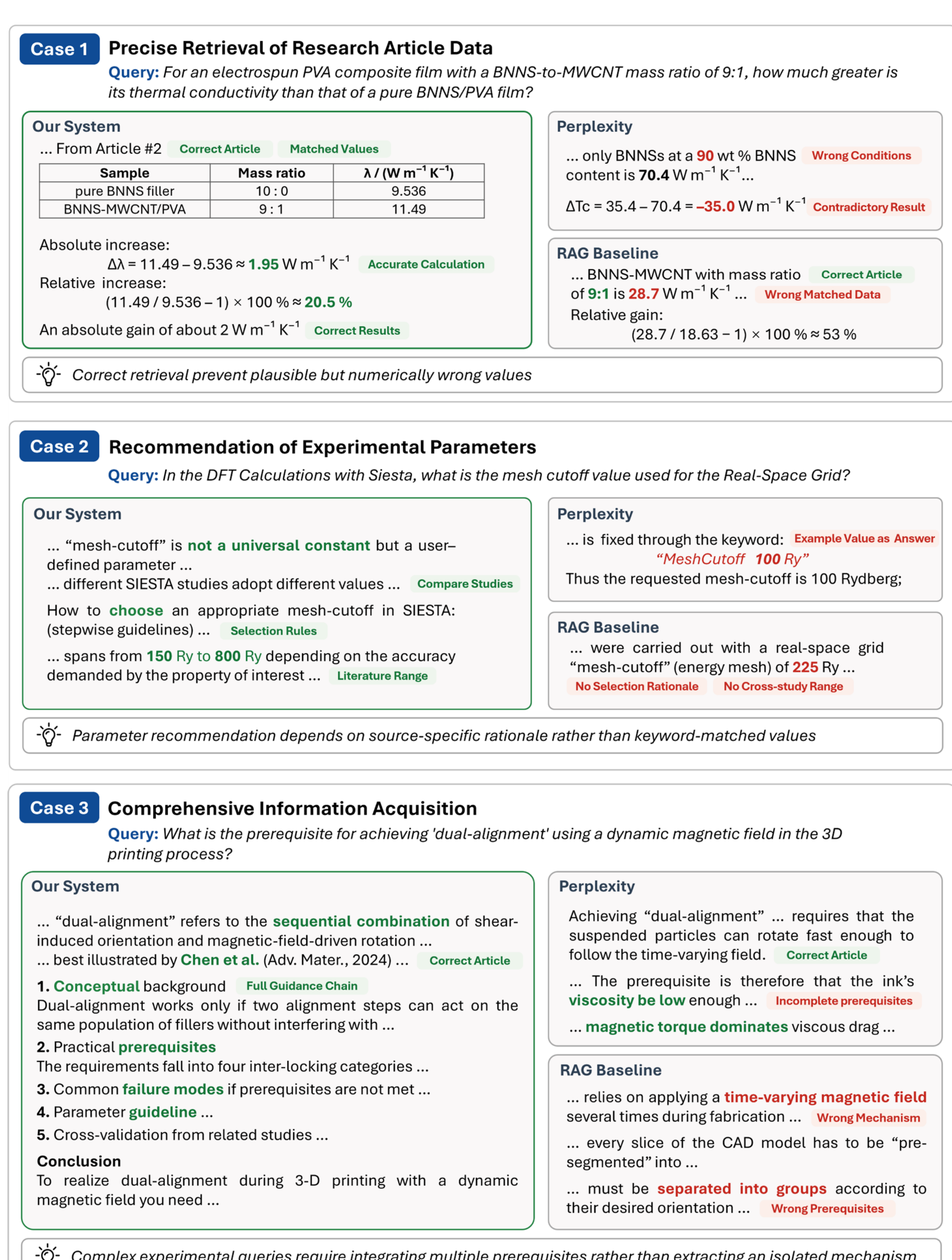

**Figure 5.** Question-answering performance of a language-native heterogeneous knowledge base across three tasks of increasing complexity: (i) a factual query, (ii) multi-step reasoning with rule induction, and (iii) counterfactual analysis of failure modes. Baselines—perplexity-based retrieval and web-augmented ChatGPT—suffer from misinterpretation, incorrect citations, factual hallucinations, and shallow reasoning. In contrast, our system accurately retrieves relevant literature, extracts precise evidence, and demonstrates advanced scientific reasoning, including induction (Case 2) and structured counterfactuals (Case 3).

**Case 3 (Mechanistic Reasoning):** The query requests an explanation of the “dual-alignment” mechanism of the thermal conducting filler and its prerequisites. Our framework decomposes the process into sequential physical steps—shear-induced axial orientation followed by magnetic-field-driven rotation—and reconstructs the associated causal chain. It further identifies key constraints, yielding a set of experimentally actionable conditions. By comparison, the baseline RAG exhibits semantic drift. Failing to capture the dependency of the process, it conflates this concept specific magneto-shear coupling with “multi-directional orientation” from unrelated 3D printing literature, producing an explanation inconsistent with the underlying physics.

These cases underscore a critical insight that when other pipeline components are held constant, organizing domain knowledge into lightly structured, self-contained modules better preserves context than chunk-based indexing. Each module encapsulates the full narrative arc—from abstract to experimental conditions to results to conclusion—do not depend heavily on neighboring context. This contextual integrity ensures that the reasoning chains required for multi-hop inference[63] remain intact.

In contrast, chunk-based evidence is inherently fragmented and short-range. When multiple chunks are aggregated, often out of order or drawn from different contexts, contextual inconsistency may distort the original meaning and mislead downstream reasoning. This structural fidelity is what empowers our system to deliver not just accurate facts, but verifiable, expert-aligned guidance for BNNS-related inquiries.

## Experience-Guided Iterative Optimization of BNNS Exfoliation

### The Experience-Augmented Iterative Design Protocol

Extending the framework's capabilities beyond retrieval-grounded recommendation, we demonstrate its use for iterative, experience-guided optimization in experimental design, leveraging the heterogeneous database to integrate empirical narratives. This addresses the trial-and-error character of materials research by enabling LLMs to refine protocols dynamically on the basis of preserved linguistic descriptions of mechanisms, process conditions and prior outcomes[39,43]. In a representative implementation, retrieval from the database yields lightly structured records from more than 200 articles on BNNS preparation by ball milling.. These records are iteratively distilled and updated into a comprehensive guide that serves as an external reference for LLMs LLM-assisted design of BNNS preparation processes.

Starting from a user objective (e.g., “BNNS ball milling—parameters, process, characterization”), the system queries the heterogeneous database and feeds retrieved modules into an iterative loop. An experience (E) template is initialized by summarization and distillation, followed by prompt-guided evolution (see **Supplementary Data 6** for experience-update protocols), batch integration of new chunks, and optional human-in-the-loop quality control (entity checks, hallucination screening). The resulting experience base provides an enhanced context for the LLM, enabling it to generate coherent, literature-grounded design guidance rather than relying on fragmented retrieval alone.

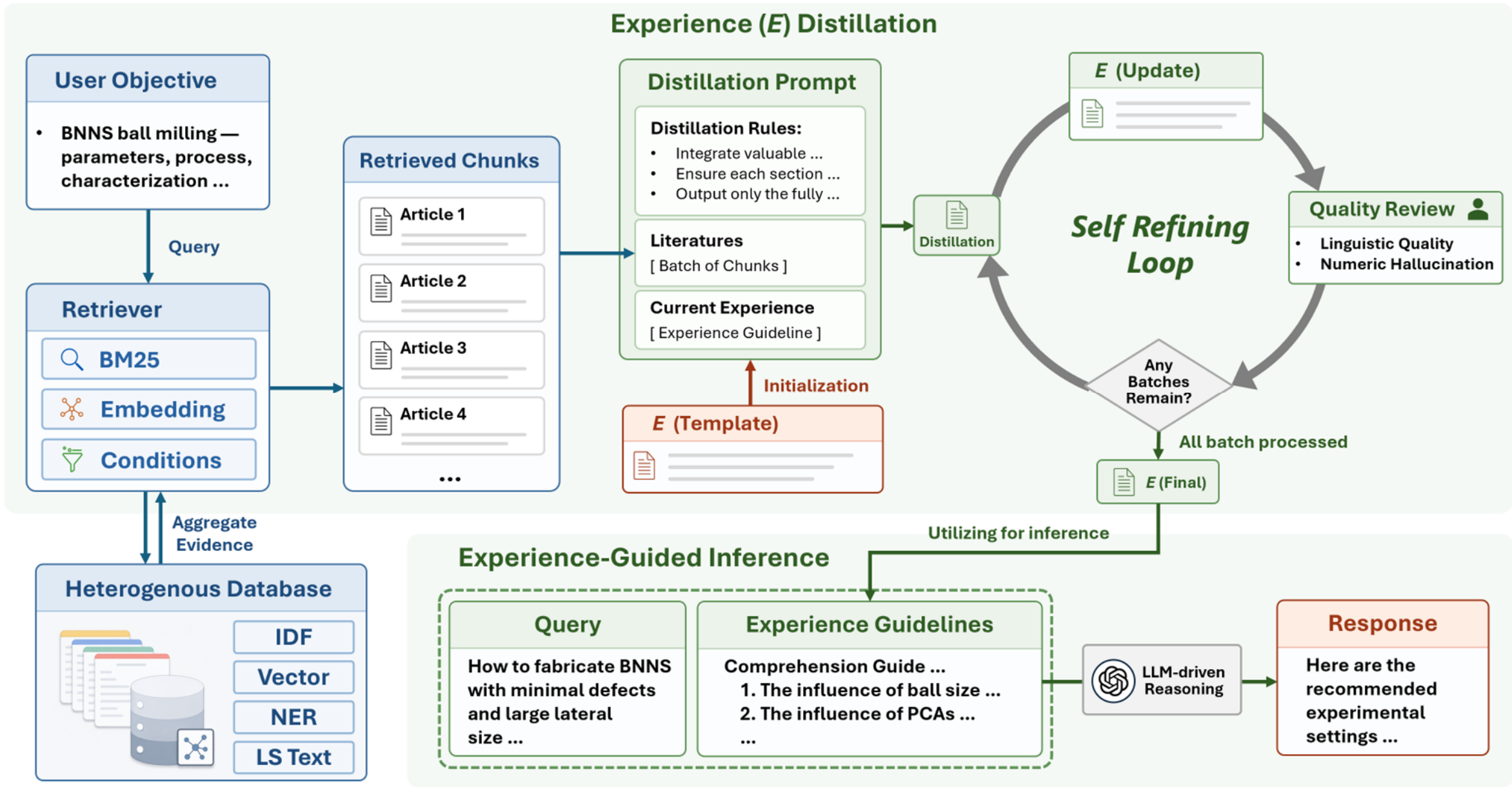

**Figure 6**: Schematic of the experience-augmented reasoning system architecture. A user objective is first defined, which guides the retrieval of relevant information from heterogeneous databases. The retrieved results are then processed through an iterative LLM-driven enhancement loop consisting of: (i) experience summarization and integration by the LLM; (ii) quality assessment and hallucination analysis; (iii) refinement and re-integration of experiences; and (iv) iteration back to step 1. or finalization of an enhanced knowledge base, which serves as an experience context for LLM inference.

### Experimental Realization of Controlled Exfoliation

To evaluate the framework in a realistic R&D scenario, we conducted a case study commissioned by a lithium battery manufacturer (facilitated by Shanghai Boronmatrix) targeting BNNS fillers for gel electrolytes. This application imposes stringent and partly conflicting specifications, where the nanosheets must be ultrathin (<20 nm), maintain controlled lateral dimensions, provide high specific surface area for ion transport, and carry hydrophilic functional groups (e.g., hydroxyl) for salt dissociation and dispersion, all under tight purity constraints. These requirements define a representative challenge presents a processing, in which success depends not only on the choice of exfoliation route, but also on how that route is adapted to local experimental conditions. Although numerous BNNS exfoliation methods have been reported, direct replication often fails because key operating conditions are only partially specified or are not directly transferable across laboratories, for example because of differences in milling energy input, vessel geometry or post-treatment conditions. The main challenge is therefore not simply to identify a literature precedent, but to iteratively adapt and refine candidate protocols so that they function reliably under a given set of local constraints.. We utilized this scenario to examine whether the framework could accelerate that process by linking literature evidence with iterative, feedback-informed design. Guided by the "Experience Guideline" (**Supplementary Data 6**), the LLM (o3 pro) supported a three-stage iterative optimization for BNNS exfoliation. The system analyzed the outcome of each experiment, interpreted the results in light of the accumulated experience context, and revised the working hypothesis and process parameters for the subsequent round. **Figure 7** (and **Supplementary Table S3** for comprehensive parameter sets) summarizes this optimization trajectory by contrasting the experimentally implemented parameters with the model's strategic rationale at each stage.

**Round 1: Exploration.**

In the first round, the system proposed a decoupled strategy consisting of mechanical exfoliation in an IPA/water medium followed by a separate urea functionalization step. This design was intended to maximize controllability and allow intermediate characterization. To compensate for the low viscosity of the IPA/water system, 1 wt% PVP was introduced to enhance effective shear during milling.

Experimentally, however, this strategy yielded predominantly thick flakes, with an average thickness of ~200 nm, together with noticeable media wear and persistent PVP residue. These results indicated that the initial route was insufficient for achieving efficient exfoliation under the target constraints.

**Mechanism Pivot from Impact-Dominated to Shear-Dominated Processing.**

Based on the outcome of Round 1, the design shifted from a low-viscosity, partially impact-dominated regime toward a more viscous and shear-dominated strategy. The separate urea functionalization step was therefore integrated into a single-step urea-paste milling protocol, in which the high viscosity of the paste was expected to improve shear transmission during exfoliation. To implement this shift, the milling media were reconfigured to a 5/2/1 mm mixed-ball assembly at a ball-to-material ratio of 100:1. This adjustment substantially reduced the nanosheet thickness to ~40 nm, but the resulting product still showed evidence of bead wear and residual organic contamination, indicating that further refinement was needed.

**Round 3: Refinement.**

In the third round, the protocol was further refined in response to the residual contamination and wear observed after Round 2. PVP was eliminated, on the basis that the urea paste itself provided sufficient viscosity for stabilization, and the milling media were adjusted to a 2 mm/1 mm microbead combination. This change was designed to increase shear density while reducing impact energy below the threshold associated with ceramic wear. Under these conditions, the system converged on a final protocol that yielded BNNS with an average thickness of ~6.5 nm, demonstrating that iterative revision of both strategy and process parameters was necessary to meet the target specifications.

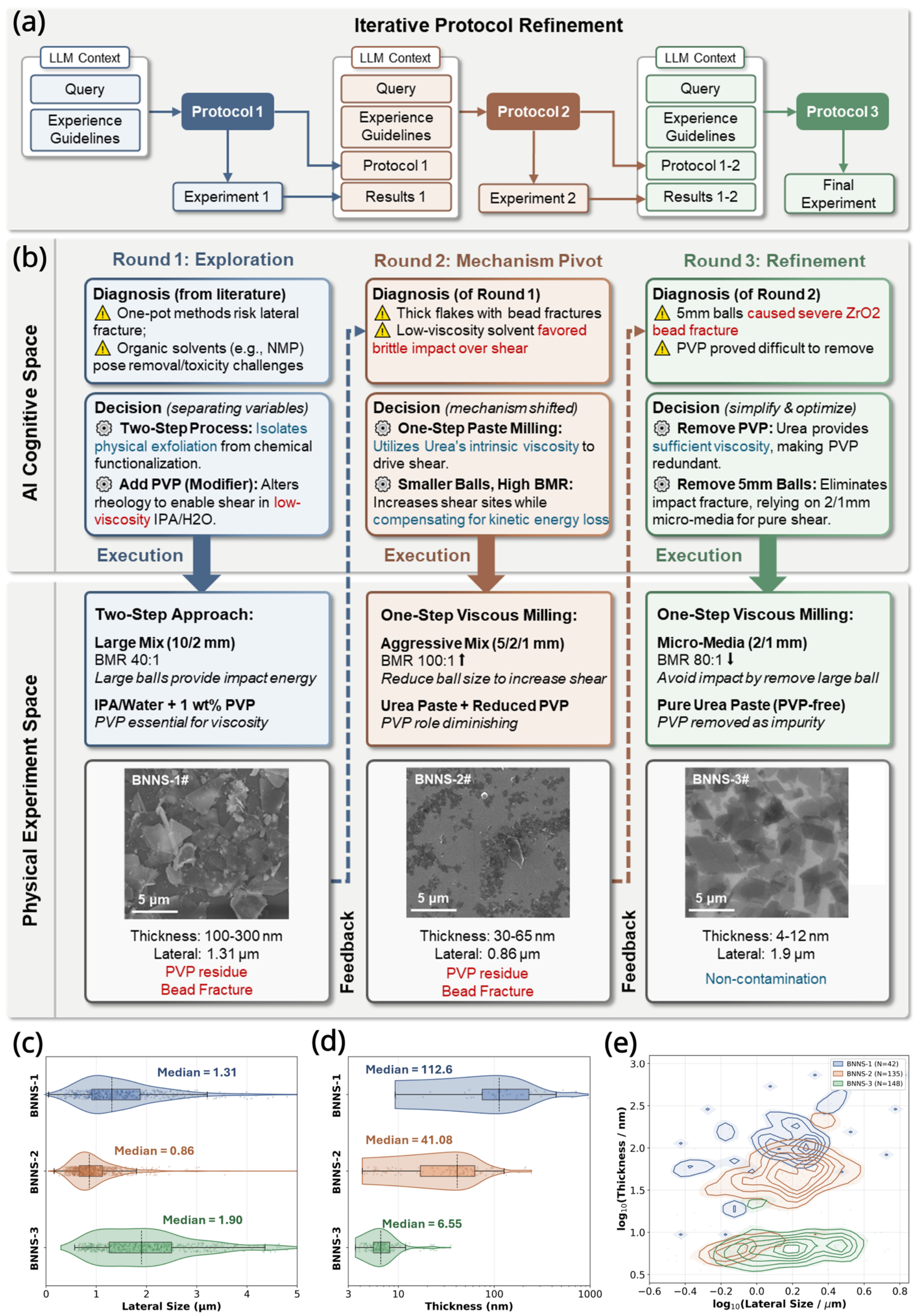

**Figure 7:** LLM-driven closed-loop optimization of ball-milling protocols for BNNS gel-electrolyte fillers. (a) Prompt-to-protocol workflow for iterative LLM-guided protocol generation. (b) Reasoning trajectory showing how diagnosis and experimental feedback led to protocol decisions, outcome interpretation, and refinement over three rounds. (c) and (d) lateral size and thickness distributions during iterative optimization. (e) lateral size versus thickness contours showing morphological evolution across the three rounds.

**Outcome and Efficiency Analysis:** This iterative design trajectory converged on a protocol yielding BNNS with an average thickness of ~6.5 nm—a >90% reduction compared to the initial trial (experimental characterization results are provided in **Supplementary Figs. S8-S13**). Supplementary characterization further supported the quality of the final product, including hydroxyl-related surface functionality in XPS, improved XRD signatures, and enhanced dispersion stability, consistent with the target requirements of the application (Supplementary Figs. S8–S13).

While the "urea-paste" methodology draws upon the seminal work of Wu et al.[64], its deployment here represents a reasoning-driven adaptation rather than direct retrieval. The final process parameters, particularly the ball-to-material ratio and the multi-scale media combination, were refined in response to the specific failure modes and local experimental constraints identified during the iterative process. Additional analyses also indicate that experience accumulation broadened the range of plausible initial designs while preserving convergence toward a consistent optimization direction in later rounds, supporting the robustness of the experience-guided workflow. Detailed comparisons are provided in **Supplementary Table X**.

This entire decision trajectory—from strategy selection to parameter refinement—was guided by the system's experience-informed reasoning. This was particularly evident in Round 2, when the framework disfavoured a conventional proposal based on viscous additives such as sucrose to enhance shear stress, a common heuristic in laboratory ball-milling. Instead, it identified several practical risks associated with this route, including cavitation effects that could impede shear transfer, heat accumulation difficulties in high-viscosity milling, and, most critically, the industrial non-viability of the requisite post-treatment (time-consuming dialysis). By prioritizing the urea-paste route—which allows for efficient water-washing purification—the system demonstrated a comprehensive foresight that transcended immediate technical feasibility to encompass downstream processability and scalability, effectively correcting the bias of human experts.

In traditional materials R&D, moving from literature screening and method selection to the refinement of process parameters can require extended trial-and-error under comparable development constraints. In this case, the framework converged on a satisfactory protocol within three focused experimental rounds under the specified laboratory conditions, showing that "Experience" functioned not merely as a static database, but as an evolving context for reasoning, protocol revision and process acceleration.

## Conclusion and Perspective

This work positions materials preparation, exemplified here by BNNS exfoliation, as a knowledge-grounded planning problem rather than a search for fixed numerical optima alone. We show that procedural knowledge recorded in text can be organized into a lightly structured and computable form that supports adaptive decision-making under path-dependent and multivariate constraints. The BNNS study provides a proof of concept that this approach can guide iterative process refinement and converge on high-quality ultrathin nanosheets. In the broader context of increasingly intelligent and automated materials research, these results highlight the importance of textual experimental knowledge as an operational foundation for AI-enabled synthesis planning.

More broadly, this study highlights the importance of knowledge representation in AI-enabled materials research. Many synthesis problems are constrained less by the lack of numerical data than by the difficulty of mobilizing procedural knowledge dispersed across literature, protocols and laboratory experience. By organizing such knowledge into lightly structured, provenance-linked modules, our framework provides a practical route for connecting human-readable experimental evidence with machine-usable reasoning. In this sense, the framework is not intended to replace structured datasets or conventional machine learning, but rather to complement them by addressing aspects of materials decision-making that depend on sequence, contingency, trade-offs and experimental feedback.

This perspective also points to a broader direction for future research. As AI systems become more deeply integrated into materials development, progress will depend not only on stronger models, but also on knowledge substrates that can support grounded reasoning over complex experimental workflows. Lightly structured text databases may therefore provide an important foundation for linking literature-derived evidence, accumulated laboratory experience and future automated execution platforms. More generally, our results suggest that language-grounded reasoning can extend AI from literature assistance toward more active roles in synthesis planning, iterative optimization and intelligent materials development.

## Acknowledgement

This work was supported by the Fundamental and Interdisciplinary Disciplines Breakthrough Plan of the Ministry of Education of China (JYB2025XDXM410), Natural Science Foundation of China (Grant No. 22573072, 62476194, U23B2049), and “the Fundamental Research Funds for the Central Universities.”